\magnification=1200

\font\bigbf=cmbx9  scaled\magstep2 \vskip 0.2in \centerline{\bigbf
Determinant of the Potts model transfer matrix} \vskip 0.1in
\centerline{\bigbf and the critical point}

\vskip 0.4in \font\bigtenrm=cmr10 scaled\magstep1
\centerline{\bigtenrm  Behrouz  Mirza } \vskip 0.2in

\centerline{\sl Department of Physics, Isfahan University of
Technology, Isfahan 84154, Iran }



\vskip 0.1in

\centerline{\sl E-mail: b.mirza@cc.iut.ac.ir}

\vskip 0.2in \centerline{\bf ABSTRACT} \vskip 0.1in

By using a
 decomposition of the transfer matrix of the $q$-state Potts Model on a three dimensional
 $ m \times n \times n $ simple cubic lattice its determinant is calculated
 exactly. By using the calculated determinants a formula is
 conjectured which approximates the critical temperature for a
 d-dimensional hypercubic lattice.

\vskip 0.1in \noindent PACS numbers: 05.50.+q, 02.70.-c

\noindent Keywords: Potts Model; Exactly Solvable Models;
Critical temperature.

\vfill\eject

\vskip 1in \centerline{I. \bf   Introduction } \vskip 0.1in

The two dimensional $q$-state Potts models [1,2] for various $q$
have been of interest as examples of different universality
classes for phase transitions and, for $q=3,4$ as models for the
adsorption of gases on certain substrates [3,4,5]. For   $q \geq
 3$ the free energy has never been calculated in closed form for
arbitrary temperature.  Some exact results have been established
for the model: from a duality relation, the critical point has
been identified [1]. The free energy and latent heat [6,7,8], and
magnetization [9] have been calculated exactly by Baxter at this
critical point, establishing that the model has a continuous,
second order transition for $q \leq 4$ and a first order
transition for  $q \geq 5$. Baxter has also shown that although
the $q = 3$ model has no phase with antiferromagnetic long-range
order at any finite temperature there is an antiferromagnetic
critical point at  $T = 0$ [9]. The values of the critical
exponents (for the range of $q$ where the transition is
continuous) have been determined [10-12]. Further insight into the
critical behaviour was gained using the methods of conformal
field theory [13]. Our knowledge about the three dimensional
Potts model is much less than the two dimensional case. The three
dimensional 3-state Potts ferromagnet serves as an important model
in both condensed matter as well as high energy physics [2].
Experimental realizations are structural in some crystals, and
theoretically this model has attracted much interest as a simple
effective model of finite-temperature pure-gauge QCD.
Consequently it has been studied in the past few years by many
authors using quite a variety of different techniques, for an
incomplete list see [14-20, and refs therein]. By using Monte
Carlo simulations with the help of standard finite size scaling
methods the characteristic parameters of the phase transition
(transition temperature, latent heat, etc) have been estimated
with varying accuracy [21]. As a result there is by now general
consensus that this model undergoes on a simple cubic lattice a
weak first order phase transition from a three fold degenerate
ordered low temperature phase to a disordered phase at high
temperature. However, our knowledge about exact value of the
partition function and spectrum of the three dimensional q-state
Potts model is almost nothing. It is thus of continuing value to
obtain further information about the three dimensional Potts
model. In this paper by using edge transfer matrices the product
of the eigenvalues for the transfer matrices of the $q$-state
Potts model on the three dimensional lattices is calculated. The
paper is organized as follows: In section II, by using the
standard representation of the transfer matrices for the
$q$-state Potts model on
 $2 \times n \times n$ lattices and using a method which is suggested
 by R.J.Baxter
 [29],
 the determinants of the transfer
matrices are calculated exactly and then the calculation is
generalized to the three dimensional $m \times n \times n$
 simple cubic lattices. Finally from the general relation for the
 determinant a formula for the $d$-dimensional $q$-state Potts model is
 conjectured which approximates the critical temperature with a few percent error.

\vskip 0.2in \centerline{II. \bf   \ Determinant of the Potts
model transfer matrix} \vskip 0.1in

 The $q$-state Potts model has served as a valuable model in the
 study of phase transition and critical phenomena. On a lattice,
 or more generally on a graph $G$, at temperature $T$ this model is
 defined by the partition function:
$$
Z(G,q,k)=\sum_{\{\sigma_n\}} e^{-\beta\, H} \eqno(1)$$ \noindent

\noindent  with the Hamiltonian

$$H=-J\sum_{<i,j>} \delta(s_i, s_j) \eqno(2)$$

\noindent  where $\delta(s_i, s_j)$ is the kronecker delta and
$s_i=1,\ldots ,q$ are the spin variables on each vertex $i\in G$
, $\beta=(k_B T)^{-1}$, $k=\beta J $; and $<i,j>$ denotes pairs of
adjacent vertices. Consider an  $n \times n$ square lattice with
periodic boundary conditions. For the two dimensional Ising model,
partition function can be written as trace of the product of
transfer matrices and the eigenvalues can be calculated exactly
[22-27]. There are several representations for the transfer
matrix of the two dimensional $q$-state Potts model [22,28] which
can be used to represent the transfer matrices for m-layer
models. Determinant of the transfer matrix for the two
dimensional Potts model can be calculated by different methods
[29,30,31]. It is quite hard if not impossible to generalize our
method [30] to higher dimensions (d $> 2$ ). We will use a method
which is suggested by R.J.Baxter and generalize it to calculate
the determinant of the transfer matrix for the q-state Potts
model on three dimensional $m \times n \times n $ simple cubic
lattices. At first we calculate the determinant for the two-layer
lattices, generalization to m-layers is simple and
straightforward. For a two layer lattice the transfer matrix
element connecting the stripes $ s_1,\ldots , s_n, t_1,\ldots
,t_n $ and $s^\prime_1,\ldots ,s^\prime_n ,t^\prime_1 ,\ldots ,
t^\prime_n $ can be written as

$$< s_1,\ldots , s_n, t_1,\ldots ,t_n \mid P \mid s^\prime_1,\ldots ,
s^\prime_n ,t^\prime_1 ,\ldots , t^\prime_n >= $$

$$  \prod_{k=1}^n e^{ {k  [\delta(s_k, s_{k+1})+  \delta(t_k , t_{k+1}) +\delta(s_k,t_k)] }} \  e^{  {k
[\delta(s_k, s_k^\prime) +\delta(t_k, t_k^\prime)]}} \eqno(3) $$

\noindent We consider the stripe $ s_1,\ldots , s_n, t_1,\ldots
,t_n $ in a plane with a periodic boundary condition along one
direction $(x)$ and  free boundary conditions in the other
direction $(y)$ , i.e. $ s_{n+1}=s_1, t_{n+1}=t_1 $. The transfer
matrix operates along the z direction which is perpendicular to
the $x-y$ plane and connects the two stripes.  \noindent Let us
define two $q^{2n} \times q^{2n} $  matrices $V_2 $ and
$V^\prime_1 $ whose matrix elements are given by

$$ <s_1, \ldots  ,t_n  \mid V_2 \mid s_1^\prime , \ldots ,
  t^\prime_n > \equiv  \delta_{s , s^\prime} \ \delta_{t , t^\prime} \prod_{k=1}^n e^{ k
[\delta (s_k,s_{k+1}) +\delta (t_k,t_{k+1})+ \delta (s_k,t_{k})]}
\eqno(4) $$

$$ <s_1, \ldots  ,t_n  \mid V_1^\prime \mid s_1^\prime , \ldots
 ,  t^\prime_n> \equiv \prod _{k=1}^n
e^{k [\delta (s_k,s_k^\prime)+ \delta (t_k,t_k^\prime)]} \eqno(5)
$$

\noindent where

$$\delta_{s , s^\prime}=\prod_{i=1}^n \delta(s_i,s_i^\prime) \ \  \ \ \ \ , \ \ \ \
\delta_{t , t^\prime}=\prod_{i=1}^n \delta(t_i,t_i^\prime)
\eqno(6) $$

\noindent Both $V_2 $ and $V_1^\prime $ are products of n
matrices [29]:

$$ V_2 = W_{12}W_{23}\ldots W_{n,1} \eqno(7) $$

$$ V_1^\prime = U_1 U_2 \ldots U_n \eqno(8) $$

\noindent where $W_{\nu,\nu+1}$ is a diagonal matrix with entries

$$ [W_{i,i+1}]_{st,s^\prime t^\prime }= e^{k [\delta(s_i,s_{i+1})+ \delta( t_i,t_{i+1})+\delta(s_i,t_i)]}
\delta_{s , s^\prime} \delta_{t , t^\prime} \eqno(9) $$

\noindent and $U_i\   (i=1,\ldots , n)$ is the matrix with entries

$$ [U_{i}]_{st,s^\prime t^\prime }= e^{k [\delta(s_i,s^\prime_{i})+ \delta( t_i ,t^\prime_ {i})]}
\prod_{j=1,j\neq i}^n \delta(s_j,s_j^\prime)
\delta(t_j,t_j^\prime)\eqno(10) $$

\noindent  Simultaneously permuting $s_1,\ldots,t_n$ and
$s_1^\prime,\ldots , t_n^\prime$ is equivalent to a rearrangement
of the rows and columns of these matrices, so

$$ \det W_{12}=\det W_{23}=\cdots= \det W_{n1} \eqno(11)$$

$$ \det U_{1}=\det U_{2}=\cdots= \det U_{n} \eqno(12)$$

\noindent Hence

$$ \det P=(\det W_{12} \det U_1)^n \eqno(13)$$

\noindent $W_{12}$  is a diagonal matrix and its determinant is
the product of its diagonal entries

$$\eqalignno{\det W_{12} &= \prod_{s_1,\ldots,s_n} \prod_{t_1 , \ldots , t_n}
e^{k\delta(s_1,s_2)} e^{k \delta(t_1,t_2)} e^{k \delta(s_1,t_1)}
&(14)\cr  &= \bigg[  \prod_{s_1,s_2}\prod_{t_1,t_2}
e^{k\delta(s_1,s_2)}e^{k\delta(t_1,t_2)}e^{k\delta(s_1,t_1)}
\bigg]^{q^{2n-4}} &(15)\cr
     &=  \bigg[ e^{3 k q^3} \bigg]^{q^{2n-4}}      &(16)\cr
 &= e^{3k q^{2n-1}} &(17)\cr }$$

\noindent $U_1$ is a block-diagonal matrix, consisting of
$q^{2n-2}$ diagonal blocks, each of the form

$$ A= (e^k I_{q \times q }+ \sigma_{q\times q} )\otimes (e^k I_{q \times q }+ \sigma_{q\times q}
)\eqno(18)
$$

\noindent where $\sigma $ is a $q\times q $ matrix with zero
diagonal elements and unit elements on all other entries and $I $
is a $q\times q$ unit matrix.  The determinant of the $q^2\times
q^2 $ matrix $A$ is given by,

$$ \det A = (e^k+q-1)^{2q}\ (e^k -1)^{2q(q-1)} \eqno(19) $$

\noindent by using the following definitions

$$ x={{e^k+q-1}\over{e^k-1}} \ \ \ \ , \ \ \ \ y=e^k \eqno(20) $$

\noindent we arrive at

$$ \det U_1=(\det A)^{q^{2n-2}}=[x^{2q} (y-1)^{2q^2}]^{q^{2n-2}}  \eqno(21) $$

\noindent so the determinant of the transfer matrix is given by

$$ \det P= [ x(y-1)^q ]^{2n q^{2n-1}}y^{3n q^{2n-1}} \eqno (22) $$

\noindent For $m$-layer lattices after a straightforward
calculation we get to the following results. The matrix $A$ is a
direct product of $m$ factors

$$ A= (e^k I_{q \times q }+ \sigma_{q\times q} )\otimes (e^k I_{q \times q }+ \sigma_{q\times q}
)\otimes \cdots \otimes (e^k I_{q \times q }+ \sigma_{q\times q}
)\eqno(23)
$$

\noindent and so we arrive at

$$ \det A= x^{mq^{m-1}}(y-1)^{m q^m} \eqno(24)$$

$$ (\det U_1)^n= (\det A)^{n q^{m(n-1)} } \eqno(25) $$

\noindent $W_{12}$  is again a diagonal matrix and its determinant
is the product of its diagonal entries

$$ \eqalignno{\det W_{12} &= \prod_{s_1^{(1)},\ldots,s_n^{(1)}} \cdots \prod_{s_1^{(m)} , \ldots ,
s_n^{(m)}} e^{k \sum_{i=1}^m \delta(s_1^{(i)},s_2^{(i)})} \  e^{k
\sum_{i=1}^{m-1} \delta(s_1^{(i)},s_1^{(i+1)})} &(26)\cr  &=
\bigg[ \prod_{s_1^{(1)},s_2^{(1)}} \cdots
\prod_{s_1^{(m)},s_2^{(m)}} e^{k \sum_{i=1}^m
\delta(s_1^{(i)},s_2^{(i)})}\ e^{k \sum_{i=1}^{m-1}
\delta(s_1^{(i)},s_1^{(i+1)})} \bigg]^{q^{m(n-2)}} &(27)\cr
     &=  \bigg[ e^{(2m-1) k q^{2m-1}} \bigg]^{q^{m(n-2)}}      &(28)\cr
 &= e^{(2m-1) k q^{mn-1}} &(29)\cr }$$

\noindent Finally we arrive at the following formula for the
determinant of the transfer matrix for the $q$-state Potts model
on a three dimensional $m\times n\times n$ simple cubic lattice

$$\det P = [x (y-1)^q]^{mn q^{[mn-1]}}y^{n(2m-1)q^{[mn-1]}} \eqno(30) $$

\noindent Considering a periodic boundary condition in both x and
y directions (a plane which is perpendicular to the direction
that transfer matrix operates), a straightforward calculation
shows that the determinant is given by the following simple
equation

$$\det P = [x y^2 (y-1)^q]^{m^2 q^{m^2-1}} \eqno(31) $$

\noindent There is a temperature where the product of the
eigenvalues does not depend on the size of lattices. It should be
noted that all of the eigenvalues are positive and  $\det P=1$ can
be simplified to the following equation

$$ x y^2 (y-1)^q =1 \eqno(32) $$

\noindent It is clear that (29) has a solution for $ k= \beta J $
 which is a function of $q$ and does not depend on $m$ or the size
 of lattices. There is another interesting point about the
 periodic boundary conditions in all directions. Consider a one
 dimensional lattice with a periodic boundary condition, the transfer matrix is given by

 $$P^{(1)}= e^k I_{q\times q}+\sigma_{q\times q} \eqno(33) $$

 \noindent The determinant can be calculated easily

 $$\det P^{(1)} =x(y-1)^q \eqno(34) $$

 \noindent For a two dimensional $m\times m $ lattice with a periodic boundary
 condition the determinant is given by [29,30,32]

 $$\det P^{(2)} =[x y (y-1)^q]^{m q^{m-1}} \eqno(35) $$

 \noindent and the determinant for a three dimensional  $m\times m \times m$  lattice with a periodic boundary condition
 for the two dimensional plane which is perpendicular to the direction that the transfer matrix operates is
 given by

$$\det P^{(3)} = [x y^2 (y-1)^q]^{m^2 q^{m^2-1}} \eqno(36) $$

\noindent This calculation can be extended to the $d $
dimensional lattices. Considering a toroidal boundary condition
for the $d-1$ dimensional hypersurface which is perpendicular to
the direction that the transfer matrix operates we arrive at the
following relations

$$ (\det W_{12})^m = y^{s(d-1) q^{s-1 }} \eqno(37)$$

$$ (\det U_1)^m = (\det A)^{m q^{(s/m)(m-1)}} \eqno(38)$$

$$ \det A = x^{({s/ m})q^{(s/m)-1}}(y-1)^{(s/m)q^{(s/m)}} \eqno(39) $$

\noindent where $s=m^{d-1} $. Finally we arrive at the
 following general formula for the determinant at any
 dimension $(d=1,2,3,\ldots)$
 for $m^{(1)}\times \cdots \times m^{(d)}$ hypercubic lattices

$$ \det P^{(d)} =[x y^{d-1} (y-1)^q]^{s q^{s-1}} \eqno(40) $$

\noindent For a two dimensional lattice the critical point
corresponds to the maximum of the following factor of the
determinant.

 $$ {xy(y-1)^q \over (y-1)^q}=xy \eqno(41) $$

\noindent We may expect that the maximum of the following factor
of the d-dimensional determinant will give us an approximate
formula for the critical temperature (we cannot prove this except
for the two dimensional lattices).

$$ {xy^{d-1}(y-1)^q \over (y-1)^q}=xy^{d-1} \eqno(42) $$

\noindent The maximum occurs at the following point

$$ k_{max}(q,d)= \rm {ln}\ [ {{2(q-1)-d(q-2)+  \sqrt{4 \ q \ (d-1) +q^2 (d-2)^2 }}\over {2(d-1)}} \ ]
\eqno(43) $$

\noindent So we may conjecture that (43) is an approximation for
the critical temperature. Interested reader may compare (43) with
data which is given in [2] (see tables II and III in page 256)
and see that it is indeed a good approximation for the critical
temperatures.

 $$ k_{crit}(q,d)\approx k_{max}(q,d) \eqno(44)$$

\noindent It may be interesting to extend these results to other
lattices with different boundary conditions. This decomposition
of the transfer matrix may be useful for obtaining other exact
results for the $m$-layer or three dimensional lattices.

 \vskip 0.2in
\centerline{V. \bf Conclusion} \vskip 0.1in

In this work determinant of the transfer matrix for the $q$-state
Potts model on a three dimensional $m\times n\times n$ simple
cubic lattice is calculated exactly and the critical temperature
is approximated by a conjectured formula.

\vskip 0.2in \centerline{\bf \ \ Acknowledgements} \vskip 0.2in

The Author would like to thank R.J. Baxter for useful
 suggestions. I would like to thank the Isfahan University of
Technology for financial support.

\vskip 0.2in \centerline{\bf \ \  References} \vskip 0.1in

\noindent [1] \ R. B. Potts, Proc. Camb. Phil. Soc. {\bf 48}, 106
(1952).

\noindent [2] \ F. Y. Wu, Rev. Mod. Phys. {\bf 54} (1982) 235.

\noindent [3] \ S. Alexander, Phys. Lett. {\bf A54}, 353 (1975).

\noindent [4] \ A. N. Berker, S. Oslund, and F.Putnam, Phys. Rev.
{\bf B17}, 3650 (1978).

\noindent [5] \ E. Domany et al, Phys. Rev. {\bf B18}, 2209
(1978).

\noindent [6] \ R. J. Baxter, J. Phys. C {\bf 6} L445 (1973).

\noindent [7] \ R. J. Baxter et al, Proc. Roy. Soc. London, Ser.
A {\bf 358}, 535 (1978).

\noindent [8] \ R. J. Baxter, J. Stat. Phys. {\bf 28},1 (1982).

\noindent [9] \ R. J. Baxter, Proc. Roy. Soc. London, Ser. A {\bf
383}, 43 (1982).

\noindent [10] \ M. P. M. den Nijis, J. Phys A {\bf 12}, 1825
(1979); Phys. Rev. {\bf B 27},(1983) 1674.

\noindent [11] \ J. L. Black and V. J. Emery, Phys. Rev. {\bf
B23}, 429 (1981).

\noindent [12] \ B. Nienhuis, J. Appl. Phys. {\bf 15}, 199 (1982).

\noindent [13] \ V. S. Dotsenko, Nucl. Phys. {\bf B235}, 54
(1984), and refs therein.

\noindent [14] \ F. Karsch and A. Patkos, Nucl. Phys. {\bf B350}
(1991) 563.

\noindent [15] A. Yamagata, J. Phys. {\bf A26} (1993) 2091.

\noindent [16] O.F. De Alcantara Bonfim, J. Stat. Phys. {\bf
1991}.

\noindent [17] J. Lee and J.H. Kosterlitz, Phys. Rev. {\bf B43}
(1991) 1268.

\noindent [18] J.D. Wang and C.DeTar, Phys. Rev. {\bf D47} (1993)
4091.

\noindent [19] M. Schmidt, Z. Phys. {\bf B95} (1994) 327.

\noindent [20] P.Provero and S. Vinti, Physica {\bf A211} (1994)
436.

\noindent [21] H.J. Herrmann, W. Janke and F. Kasch  (eds.),
Dynamics of first order

\noindent \ \ \ \ \ \   phase transitions, World Scientific,
Singapore, 1992.

\noindent [22] \ R. J. Baxter, Exactly Solved Models in St. Phys.
(Academic Press, 1982).

\noindent [23] \ K. Huang, Statistical Physics, 2nd ed., Wiley,
New York 1987 (Chapter 15).

\noindent [24] \ B. Bergersen and M. Plischke, Equilibrium
Statistical Physics, 2nd edition,

\noindent \ \ \ \ \ \ \ World Scientific, Singapore, 1994 (Chapter
5).

\noindent [25] \  L. Onsager, Phys. Rev. {\bf 65}, 117-49 (1944).

\noindent [26] \  B. Kaufmann, Phys. Rev. {\bf 76}, 1232-1243
(1949).

\noindent [27] \  T. Shultz, D. Mattis  and E. Lieb,  Rev. Mod.
Phys. {\bf 36}, 856 (1964).

\noindent [28] \ P. Martin, Potts Models and related Problems in
Statistical Physics, World Scientific,

\noindent \ \ \ \ \ \ \ Singapore, 1991.

\noindent [29] \ R.J. Baxter, unpublished.

\noindent [30] \ B. Mirza and M.R. Bakhtiari, Physica A {\bf 343}
(2004) 311-316. cond-mat/0406501.

\noindent [31] \ S. Chang and R. Shrock, cond-mat/0404524.

\noindent [32] \ S. Chang and R. Schrok, Physica  A {\bf  296},
234-288 (2001), cond-mat/0011503.

 \vfill\eject

\bye